\documentclass[twocolumn,times]{aastex631}

\usepackage{tabularx}  
\usepackage{amsmath}
\usepackage{multirow}
\usepackage{xcolor}
    \definecolor{lmugreen}{rgb}{0.01,0.58,0.25}
    \definecolor{myorange}{rgb}{1.,0.35,0.}
    \definecolor{mygrey}{rgb}{0.5,0.5,0.5}
\usepackage{listings}
\usepackage[capitalise]{cleveref}
	\crefname{equation}{Equation}{Equations}
	\crefname{figure}{Figure}{Figures}
	\crefname{table}{Table}{Tables}
	\crefname{section}{Section}{Sections}
	
	\newcommand{\crefalt}[1]{\namecref{#1}~\ref{#1}}
	
\makeatletter
\usepackage{etoolbox}
\patchcmd\H@refstepcounter{\protected@edef}{\protected@xdef}{}{}
\makeatother

\let\tablenum\relax
\usepackage{siunitx}
\usepackage{etoolbox}
\usepackage{booktabs}

\received{October 18, 2021}

\submitjournal{ApJ}

\shorttitle{Contour Analysis Tool}
\shortauthors{Hutchison \& Koepferl}
 \graphicspath{{./}{figures/}}

\begin{document}

\title{Contour Analysis Tool: an interactive tool for background and morphology analysis}

\correspondingauthor{Mark A. Hutchison}
\email{markahutch@gmail.com}

\author[0000-0003-4543-8711]{Mark A. Hutchison}
\affiliation{Hochschule f{\"u}r angewandte Wissenschaften M{\"u}nchen, Lothstra{\ss}e 34, 80335 M{\"u}nchen, Germany}
\affiliation{Universit\"ats-Sternwarte, Fakult\"at f\"ur Physik, Ludwig-Maximilians-Universit\"at M\"unchen, Scheinerstr. 1, 81679 M\"unchen, Germany}

\author[0000-0003-1256-9766]{Christine M. Koepferl}
\affiliation{Hochschule f{\"u}r angewandte Wissenschaften M{\"u}nchen, Lothstra{\ss}e 34, 80335 M{\"u}nchen, Germany}
\affiliation{Universit\"ats-Sternwarte, Fakult\"at f\"ur Physik, Ludwig-Maximilians-Universit\"at M\"unchen, Scheinerstr. 1, 81679 M\"unchen, Germany}



\begin{abstract}
    We introduce the Contour Analysis Tool (CAT), a Python toolkit aimed at identifying and analyzing structural elements in density maps. CAT employs various contouring techniques, including the lowest-closed contour (LCC), linear and logarithmic Otsu thresholding, and average gradient thresholding. These contours can aid in foreground and background segmentation, providing natural limits for both, as well as edge detection and structure identification. Additionally, CAT provides image processing methods such as smoothing, background removal, and image masking. The toolkit features an interactive suite of controls designed for Jupyter environments, enabling users to promptly visualize the effects of different methods and parameters. We describe, test, and demonstrate the performance of CAT, highlighting its potential use cases. CAT is publicly available on GitHub, promoting accessibility and collaboration.
\end{abstract}



\section{Introduction}
\label{sec:introduction}

The analysis of astronomical objects is traditionally performed using 2D measurements of flux or brightness, or by combining several flux maps into extracted 2D density maps. Encoded within these maps is often physical information about the local environment (temperature, forces, chemistry, etc.). For example, histograms of column density maps or column density probability distribution functions (so-called N-PDFs) have been used extensively over the last few decades to help infer the dominant physical process(es) responsible for observed density structures in molecular clouds \citep[e.g.][]{Vazquez-Semadeni/1994,Ostriker/Stone/Gammie/2001,Vazquez-Semadeni/Garcia/2001,Ballesteros-Paredes/etal/2011,Federrath/Klessen/2013,Federrath/etal/2016,Jaupart/Chabrier/2020,Khullar/etal/2021}, typically by analysing their shape. Quiescent clouds, dominated by turbulence, generally exhibit a single lognormal N-PDF \citep{Ma/etal/2021,Ma/etal/2022}, though recent studies suggest the presence of multiple lognormal components \citep{Murase/etal/2022}. In contrast, active star-forming clouds often show a lognormal distribution with a power-law tail at higher densities, associated with gravitational contraction \citep[e.g.][]{Goodman/Pineda/Schnee/2009,Kainulainen/etal/2009,Schneider/etal/2013,Schneider/etal/2015a,Schneider/etal/2015b,Pokhrel/etal/2016,Schneider/etal/2016a,Ma/etal/2020,Wang/etal/2020,Jiao/etal/2022,Lewis/Lada/Dame/2022,Schneider/etal/2022}. Furthermore, because the mass of available molecular gas is known to be strongly correlated with the star formation rate \citep[e.g.][]{Kennicutt/1998,Bigiel/etal/2011}, N-PDFs are often used to estimate galactic star formation efficiencies \citep[e.g.][]{Handa/etal/1990,Gouliermis/etal/2010,Momose/etal/2010,Grasha/etal/2017,Fujimoto/etal/2019,Zhang/etal/2019,Villanueva/etal/2022,Maeda/etal/2023,Mattern/etal/2024}.

Although the PDF is perhaps one of the easiest quantities to measure from a density map, there are a number of challenges that have to be considered. Particularly important for observational studies is the possibility of overlapping foreground and/or background clouds along the line of sight, which disproportionately contaminate the low-density regions of the N-PDF \citep{Lombardi/Alves/Lada/2015,Schneider/etal/2015a,Schneider/etal/2015b,Ossenkopf-Okada/etal/2016}. While corrections, such as subtracting a constant offset, can mitigate these effects, there is always danger of over/under compensation \citep{Schneider/etal/2015a,Schneider/etal/2015b}. Another issues is identifying the completeness limit of the measured PDF. Open contours are, by definition, incomplete and cannot be corrected in a reliable manner. Thus, as a matter of convenience, many studies have used the last- or lowest-closed contour (LCC) as a natural boundary for clouds \citep{Lada/Lombardi/Alves/2010,Kainulainen/etal/2011,Kainulainen/Federrath/Henning/2013,Ossenkopf-Okada/etal/2016}. Later, \citet{Alves/Lombardi/Lada/2017} went one step further and classified the LCC as the natural completeness limit for a density PDF, assuming it falls above the nominal noise of the map and is not affected by an unrelated cloud within the same map. Unsurprisingly, the LCC is sensitive to windowing (e.g. instrumental or post-process masking of sources). Decreasing the field of view around a source tends to increase the value of the LCC, thereby altering the shape (and hence the physical interpretation) of the PDF within the completeness limit \citep{Kortgen/Federrath/Banerjee/2019}. The choice of level spacing or noise reduction algorithms can additionally alter the LCC. Thus, while manually identifying and verifying the LCC is straightforward in principle, the process can be tedious when optimising data parameters and viewing conditions. An automated tool would be well suited for this task.

Many tools exist that facilitate the processing and analysis of density maps in astrophysics in an automated fashion. For identifying clumps and cores, algorithms such as Clumpfind \citep{Williams/deGeus/Blitz/1994}, Gaussclumps \citep{Stutzki/Guesten/1990}, cprops \citep{Rosolowsky/Leroy/2006}, Dendrograms \citep{Rosolowsky/etal/2008}, and FellWalker \citep{Berry/2015} are widely used. CuTEx \citep{Molinari/etal/2011} provides a method based on curvature properties for source extraction. For filament identification, Filfinder \citep{Koch/Rosolowsky/2015} and DisPerSE \citep{Sousbie/2011} are notable tools. Void identification is achieved through tools like ZOBOV \citep{Neyrinck/2008} and VIDE \citep{Sutter/etal/2015}, while general tools such as Astrodendro and SCIMES \citep{Colombo/etal/2015} assist in segmentation and hierarchical decomposition of data.

Despite the diversity of available tools, none, to our knowledge, are specifically designed to identify the LCC in an image. In this paper, we introduce the Contour Analysis Tool (CAT), an interactive toolkit developed for comprehensive analysis and processing of 2D images. In addition to finding the LCC\footnote{Or, by extension, the first- or highest-open contour. In this sense, the LCC can serve as a natural \textit{cutoff} for N-PDF analysis rather than the \textit{peak} of the histogram, making it useful in various disciplines as an upper limit on background levels.}, used in N-PDF studies of star-forming regions, CAT calculates additional threshold contours that can aid in targeting, identifying, or masking regions of interest, such as high-density areas or background signals. To aid analysis, CAT is also equipped with various smoothing and background removal techniques and provides users with the ability to mask data using simple shapes, calculated contours, or custom masks that can be drawn interactively or provided by the user at runtime. CAT's versatility as both an interactive and programmatic tool, combined with its robust functionality and simple user interface, makes it an ideal tool for working with density maps.

The paper is organised as follows. In \cref{sec:methods}, we present a detailed overview of the code, covering image processing techniques, contour level calculation, and masking. \Cref{sec:testing} discusses the testing of tools that were not readily available in other packages or were integrated with other functions. Then, in \cref{sec:application}, we illustrate the performance of the code on specific use cases and conclude in \cref{sec:summary} by summarising key aspects of the code/paper. The code is publicly available on GitHub at \href{https://github.com/markahutch/ContourAnalysisTool}{https://github.com/markahutch/ContourAnalysisTool}.

\section{Methods}
\label{sec:methods}

CAT is a lightweight Python script that accepts a 2D image as a NumPy \citep{Harris/etal/2020} array and returns various critical contour levels, including the LCC, that can aid in identifying completeness limits, revealing structures, and/or estimating background levels. CAT is equipped with a number of smoothing techniques and the ability to remove low-level background noise. Interactive sessions provide flexible options to mask the data, including drawing a custom mask or using one of the calculated contours. Interactive controls update contours on the fly. Contour visibility can be toggled on and off to (i) better facilitate comparison between different background adjustments and to (ii) provide a quick visual assessment of the effect parameters have on contours without having to repeatedly rerun code. Although the current graphical user interface (GUI) only supports Jupyter environments, non-interactive features can be run in any Python 3+ environment as long as the required packages are installed (see online documentation for a complete list of requirements and dependencies). Future developments will focus on creating a GUI that can run across different environments and incorporating more advanced edge detection software. 

CAT is written in a modular form, making it both easy to use and develop. The UserInterface class provides a user-friendly interface that enables users to analyze images with little to no Python experience and minimal coding. For instance, interactive sessions can be set up with a single command\footnote{We omit optional parameters for the sake of brevity. Default parameters are used whenever optional parameters are left out and, for convenience, partial/misspelled words are also accepted as input.}, e.g.,
\begin{lstlisting}[basicstyle=\ttfamily\footnotesize\fontsize{12}{8}, language=Python]
# Set Up Interactive Instance
CAT = ContourAnalysisTool.UserInterface(image, extent, interactive=True)
\end{lstlisting}
where {\texttt{image}} is a 2D NumPy array containing the image and $\texttt{extent} = [x_{\rm min}, x_{\rm max}, y_{\rm min}, y_{\rm max}]$ is the physical extent of the image in the $x$ and $y$ dimensions, respectively. Likewise, almost all non-interactive features can be accessed via three callable functions:
\begin{lstlisting}[basicstyle=\ttfamily\footnotesize\fontsize{12}{8}, language=Python]
# Raw Image Contours
levels_raw, image_raw, plot_raw = CAT.find_contours_raw()
# Smoothed Image Contours
levels_sm, image_sm, plot_sm = CAT.find_contours_smoothed()
# No Background Image Contours
levels_nobg, image_nobg, plot_nobg = CAT.find_contours_nobackground()
\end{lstlisting}
In either mode, users can process an image either in its raw form or after applying smoothing or background-removal techniques. For each of these image processing options, the code computes the following four critical contour levels: the LCC, linear and logarithmic Otsu thresholds, and a mean gradient threshold. While the UserInterface is designed to be the main access point to the code, the underlying classes, functions, and variables that constitute the core of the code can still be directly manipulated and utilized by the user if desired.

The following subsections describe the functionality and implementation of the code in further detail.

\subsection{Image processing}
\label{sec:image_processing}

In addition to working with the original raw image provided by the user, CAT has the ability to smooth or remove low-level background noise in the image prior to contour level calculations.

\subsubsection{Smoothing}
\label{sec:smoothing}

Every smoothing method offers distinct advantages/disadvantages, balancing trade-offs between data fidelity, noise reduction, and computational efficiency. To better address a wide range of data sets with diverse noise characteristics, we implemented 10 different smoothing algorithms from the literature, each with one or more smoothing parameters to control the level of smoothing. This diverse toolkit allows the user to select the most suitable smoothing technique for their unique data set. Below, we provide an overview of each smoothing method and how it is implemented.

\paragraph{Gaussian Smoothing} Applies a Gaussian kernel to the data to average surrounding points, effectively reducing noise while preserving the signal \citep[see][]{Gonzalez/Woods/2017}. This technique is implemented using the \href{https://docs.scipy.org/doc/scipy/reference/ndimage.html}{\texttt{scipy.ndimage}} package. The key parameter is the standard deviation ($\sigma$), which determines the extent of smoothing. A larger $\sigma$ leads to a broader smoothing effect.

\paragraph{Bivariate Spline Smoothing} Fits a smooth surface over the data points using spline interpolation \citep[see][]{Dierckx/1993}, implemented via \href{https://docs.scipy.org/doc/scipy/reference/interpolate.html}{\texttt{scipy.interpolate}}. The main parameter is the smoothing factor $s$, which balances the trade-off between adherence to input data and the smoothness of the resulting surface.

\paragraph{Local Regression (LOESS)} Fits local polynomials to the data, providing flexibility in smoothing diverse data patterns \citep{Cleveland/Devlin/1988}. Implemented with the \href{https://pypi.org/project/loess/}{\texttt{loess}} library, users can control the polynomial order ($\texttt{poly\_order}$) and the fraction of data points used in local fits ($\texttt{locality\_frac}$).

\paragraph{Savitzky-Golay Filtering} Applies polynomial fitting within a moving window to smooth data while preserving significant features \citep{Schafer/2011}. Using the \href{https://docs.scipy.org/doc/scipy/reference/signal.html}{\texttt{scipy.signal}} package, users can adjust the window length ($\texttt{window\_length}$), polynomial order ($\texttt{poly\_order}$), and boundary mode ($\texttt{boundary\_mode}$).

\paragraph{Moving Average Smoothing} Averages data points within a defined neighborhood to smooth the signal and reduce random noise \citep{Smith/1997}. It uses the \href{https://docs.scipy.org/doc/scipy/reference/signal.html}{\texttt{scipy.signal}} package with a circular kernel determined by the kernel radius ($\texttt{kernel\_radius}$).

\paragraph{Wiener Filtering} Minimizes the mean square error between the estimated and true signal by incorporating statistical noise characteristics \citep{Wiener/1949}. Implemented using \href{https://docs.scipy.org/doc/scipy/reference/signal.html}{\texttt{scipy.signal}}, users control the filter size ($\texttt{mysize}$) and noise estimate ($\texttt{noise}$).

\paragraph{Bilateral Filtering} Smooths the image while preserving edges by considering both spatial and intensity differences \citep{Tomasi/Manduchi/1998}. Implemented via \href{https://scikit-image.org/docs/stable/api/skimage.restoration.html}{\texttt{skimage.restoration}}, the parameters include spatial smoothing ($\sigma_{\text{spatial}}$) and range smoothing ($\sigma_{\text{range}}$).

\paragraph{Total Variation Denoising} Minimizes the total variation of the image, reducing noise while preserving edges \citep{Rudin/Osher/Fatemi/1992}. Using \href{https://scikit-image.org/docs/stable/api/skimage.restoration.html}{\texttt{skimage.restoration}}, users can adjust the weight ($\texttt{weight}$) and the number of iterations ($\texttt{n\_iter}$).

\paragraph{Anisotropic Diffusion} Smooths data while preserving edges by allowing different diffusion rates in various directions \citep{Perona/Malik/1990}. The \href{https://scikit-image.org/docs/stable/api/skimage.restoration.html}{\texttt{skimage.restoration}} package is used, with user control over the number of iterations ($\texttt{n\_iter}$), conduction coefficient ($\kappa$), and regularization parameter ($\gamma$).

\paragraph{Non-Local Means Denoising} Compares and averages similar patches across the image to reduce noise \citep{Buades/Coll/Morel/2011}. Implemented via \href{https://scikit-image.org/docs/stable/api/skimage.restoration.html}{\texttt{skimage.restoration}}, key parameters include patch size ($\texttt{patch\_size}$) and filtering strength ($h$).

\subsubsection{Background removal}
\label{sec:background_removal}

Removing background noise is essential for isolating significant features and enhancing image analysis accuracy. Our approach for background removal involves estimating and subtracting the background using morphological filtering, implemented via the \href{https://scikit-image.org/docs/stable/}{\texttt{skimage}} package \citep{vanderWalt/etal/2014}. Users can control several parameters, including the filter shape, size, and padding mode.

The filter shape influences the morphology used to estimate the background. The `\texttt{disk}' shape is advantageous for processing circular patterns, while the `\texttt{square}' shape is more suited for grid-like structures \citep[see][]{Gonzalez/Woods/2017}. The filter size affects the extent of features removed from the background; larger sizes yield a smoother background estimate but may also remove relevant image details \citep[see][]{Smith/1997}. Edge handling is managed through the padding mode, with available options being: `\texttt{reflect}', `\texttt{symmetric}', `\texttt{wrap}', `\texttt{constant}', `\texttt{linear\_ramp}', `\texttt{edge}', `\texttt{mean}', `\texttt{median}', `\texttt{minimum}', and `\texttt{maximum}' (see \href{https://numpy.org/doc/stable/reference/generated/numpy.pad.html}{\texttt{numpy.pad}}
 documentation for details about each mode).

We start by normalizing the image to a [0, 1] range and converting it to an 8-bit format for thresholding using Otsu’s method \citep{Otsu/1979}, which effectively segments background and foreground, especially for images with bimodal histograms. After padding and normalization, a rank filter is applied to the grayscale image according to the selected filter shape and size. This rank filter averages local pixel values based on the filter’s morphology to estimate the background \citep[see][]{Russ/2011}. Subtracting this estimated background from the original image yields the background-removed image, which is then clipped to ensure the pixel values remain within a realistic range. This method makes use of the filters and morphology packages in \href{https://scikit-image.org/docs/stable/}{\texttt{skimage}} \citep{vanderWalt/etal/2014} as well as the \href{https://numpy.org/doc/stable/}{\texttt{numpy}}
 package \citep{Harris/etal/2020}.

\subsection{Contour levels}
\label{sec:contour_levels}

In addition to the LCC highlighted in \cref{sec:introduction}, the code offers three additional threshold contours: linear and logarithmic Otsu thresholds, as well as a mean gradient threshold. These contours are designed to identify the boundary between foreground and background signals using different criteria. The contours range from being completely automated (Otsu and OtsuLog thresholds) to being fully customizable (Average threshold). Meanwhile, the LCC provides an automated option within the masked region chosen by the user. Smoothing and background removal also affect the contours. Below, we describe how each of the four contours is found or calculated.

\subsubsection{LCC}
\label{sec:last_closed_contour}

The LCC identifies the lowest intensity level at which the contour remains closed within the image/mask boundaries with no open contours existing above. Note that open/closed contours that exist \textit{outside} of the masked region without intersecting the mask boundaries have no impact on the LCC \textit{inside} of the mask. More specifically, the two criteria that must be met for a contour level to be considered `closed' are (i) that it must have at least one segment within the boundary and (ii) that it cannot intersect the boundary at any point. Here, `boundary' refers to the outer edge of the intersection between the mask and image. The literature lacks details on how the LCC has been determined historically, and there are no readily available tools that provide this information. While both conditions defining closed contours are straightforward to test in principle, there are nuances in their application, at least in CAT, that need to be fleshed out in more detail.

The primary difficulty we encounter when checking for closed contours arises from masking our image without altering the underlying data array. We do this to avoid repeatedly making fresh copies of the image every time the mask is updated in an interactive session. If the edges of the image are the only boundaries to consider, it is indeed almost trivial to detect open contours because one or more of its vertices are placed on the boundary itself. However, this is no longer true when using a mask, as the probability of contour vertices falling on mask edges is virtually zero. One could mask the data on the fly in the contour function, but, if the mask is removed, the contour for the full image would then have to be recomputed. Depending on the image size and the number of contour levels used to discretize the image, this could be an expensive operation and best to avoid.

To circumvent the above issues, we employ an algorithm to find the mask edges and then cast the contour vertices to the nearest grid point to see if any of these grid points overlap the mask edge. We begin by converting the mask into a boolean array, which we then dilate by an additional two cells using the \texttt{binary\_dilation} function from the \href{https://docs.scipy.org/doc/scipy/reference/ndimage.html}{\texttt{scipy.ndimage}} package \citep{Virtanen/etal/2020}. The mask edges are then isolated by subtracting the original mask from the dilated mask. This two-cell thickness prevents diagonal `holes' that would otherwise appear if a single-celled edge happened to run diagonal to the grid (see \cref{sec:testing} for a clear example of where this fails). Note, thicknesses larger than two cells should also be avoided since they run the risk of brushing contours from the outside that should be ignored. Once the mask edges are all found, we then combine these with the image boundaries (which do not require a two-cell thickness) to account for cases where the mask extends off the edge of the image. 
As a final detail, we disregard contours that do not contain at least three unique vertices, as they cannot constitute a proper closed contour in a two-dimensional space. The LCC is identified by searching through the contours from largest to smallest and selecting the final one that satisfies both closure conditions.

\subsubsection{Otsu thresholds}
\label{sec:otsu_thresholds}

Otsu thresholding \citep{Otsu/1979} computes the optimal threshold to separate the background and foreground within an image by minimizing intra-class variance, which is the sum of the variances within each class (background and foreground). This technique is particularly effective for images with bimodal histograms, where pixel intensities form two distinct peaks corresponding to background and object regions. However, even in the absence of a clear bimodal distribution, Otsu thresholding is widely used for its simplicity and effectiveness in various image segmentation tasks. As it is often more convenient to work with logarithmically scaled data when the data spans many orders of magnitudes, we also calculate the Otsu threshold for this as well (differentiated by the name `OtsuLog').

\subsubsection{Average threshold}
\label{sec:average_threshold}

The average threshold contour level is determined by analyzing the gradient magnitudes of the image and estimating a representative background intensity level based on a user-defined threshold percentage. To begin, we compute gradient magnitudes for the image using the \texttt{gaussian\_gradient\_magnitude} function from the \href{https://docs.scipy.org/doc/scipy/reference/ndimage.html}{\texttt{scipy.ndimage}} package \citep{Virtanen/etal/2020}, with a default standard deviation of unity, to capture the local intensity changes in the image. We then apply the aforementioned percentile threshold to these gradient magnitudes, below which it is assumed gradients represent the background. Gradients below this threshold are retained, while those above it, presumably source-related, are excluded. This produces a binary map that highlights potential background regions. We process this binary map using the \texttt{find\_contours} function from the \href{https://scikit-image.org/docs/stable/api/skimage.measure.html}{\texttt{skimage.measure}} module \citep{vanderWalt/etal/2014}, specifying a contour level of 0.5 to identify the boundary line that effectively delineates the detected background regions within the binary image. Finally, we compute the average intensity along this contour line in the original image. This computed average intensity represents the effective or average contour level, providing a quantitative measure of the background intensity based on the selected threshold criteria. 

\subsection{Masking}
\label{sec:masking}

As discussed in \cref{sec:last_closed_contour}, masking significantly impacts the determination of the LCC in an image. However, we do not apply the mask when calculating other thresholding contours, as this would run counter to their purpose of identifying background levels. CAT offers users four masking options. The default method produces a circular or square mask with user-specified coordinates and size. For more complex masking, users can either upload a custom mask or draw one interactively on the plot canvas, allowing real-time visualization of changes to the LCC. The brush shapes available for interactive masking include circles and squares with adjustable sizes. Users can switch between painting or erasing the mask, and a change history is maintained to enable stepping back to previous states. As a final option, users can choose to mask the image by any of the four contours calculated for the three different image types (i.e. 12 in total). This method may be preferable when trying to isolate regions of high density (e.g. regions of active star formation; see \cref{fig:lcc_contour_mask} for an example). The averaging threshold contour can be particularly useful in this regard due to its ability to be fine-tuned across the entire range of density contours.

\subsection{Interactive plot options}
\label{sec:interactive_plot_options}

\begin{figure*}
\epsscale{0.92}
\plotone{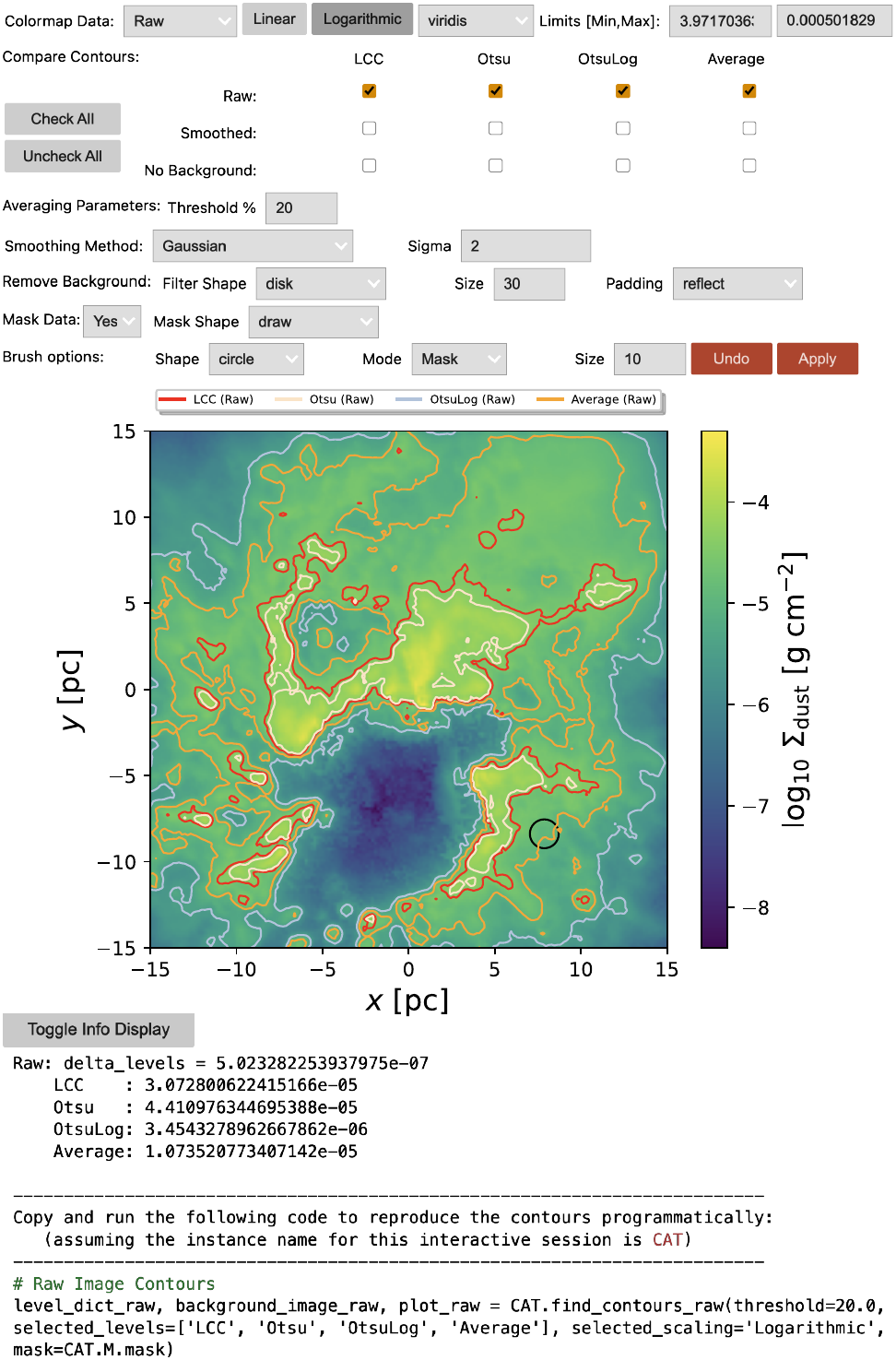}
\caption{A screenshot of an interactive session in CAT. Checkboxes control contour visibilities in the plot while dropdown menus, toggle buttons, and input fields select and modify the various methods and parameters. An additional toggle button sits below the figure and prints information values for contour levels and spacing ({\texttt{delta\_levels}}) as well as code to reproduce the results programmatically.}
\label{fig:interactive_controls}
\end{figure*}

In addition to the interactive features specific to the methods previously discussed, users have access to a small suite of interactive controls (see \cref{fig:interactive_controls}) designed to customize the figure and enhance comparisons during an interactive session. For example, the user can toggle between displaying the raw, smoothed, or background-subtracted image, with updates occurring dynamically in response to changes in the smoothing or background-subtraction parameters. Users can switch between linear and logarithmic scaling of the data to better visualize different aspects of the image. The color map and limits for the color bar can all be adjusted on the fly. Lastly, a grid of checkboxes provides control over the visibility of individual contour levels, allowing users to display or hide specific contours for any of the raw, smoothed, or background-subtracted images. Although only one image type can be displayed in the background at a time, the visibility settings for contour levels of each image type are preserved, facilitating simultaneous comparison of contour levels across different image types.




\section{Testing}
\label{sec:testing}

In large part, the code is built from an assortment of functions imported from the packages listed in \cref{sec:methods}, presumably tested by their respective developers. Thus the bulk of our testing involved (i) integrating these utilities together into a single, user-friendly interactive interface and (ii) being able to (re)produce the resulting contours and figures programmatically and providing the user with the necessary executable code. Functions for which we could find no pre-existing utility in the literature, such as the LCC, we implemented and tested as follows.

\begin{figure}
\plotone{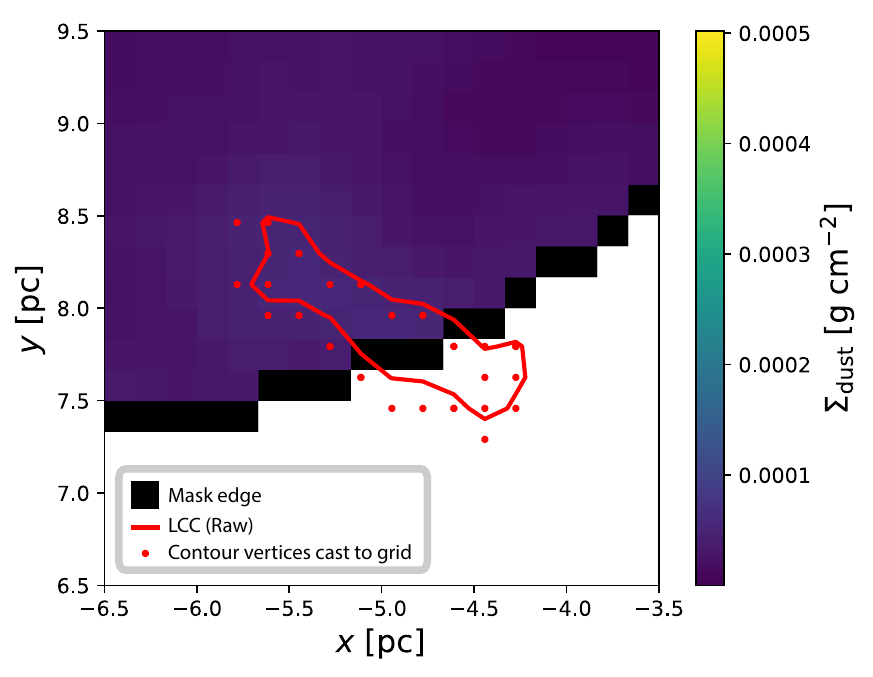}
\caption{Illustration of an intersecting contour that remains undetected due to the mask edge thickness being equivalent to a single cell width. Intersection detection involves mapping contour vertices to nearby grid points and identifying overlapping mask edges. A mask edge spanning two cells prevents diagonal gaps from forming, effectively mitigating false positives.}
\label{fig:lcc_mask_failure}
\end{figure}

Testing the algorithm to identify the LCC involved manually processing images with various windowing and masking setups. The images ranged from simple mathematical functions (e.g., dipole and sinusoidal functions, both with and without random noise) to realistic synthetic maps \citep{Koepferl/etal/2017,Koepferl/Robitaille/2017,Koepferl/Robitaille/Dale/2017a} of a star forming region taken from a smoothed particle hydrodynamics simulation from \citet{Dale/etal/2014}. For each image, we ensured that the algorithm's identified LCC met all the criteria defined in \cref{sec:last_closed_contour} while the next lowest contour did not. During testing, we determined that the mask edges must be two cells thick. A single-cell border can create diagonal gaps through which contours might pass unnoticed (see \cref{fig:lcc_mask_failure}), while a border of three or more cells risks incorrectly flagging external, non-intersecting contours near the mask edge.

\begin{figure*}  
\plotone{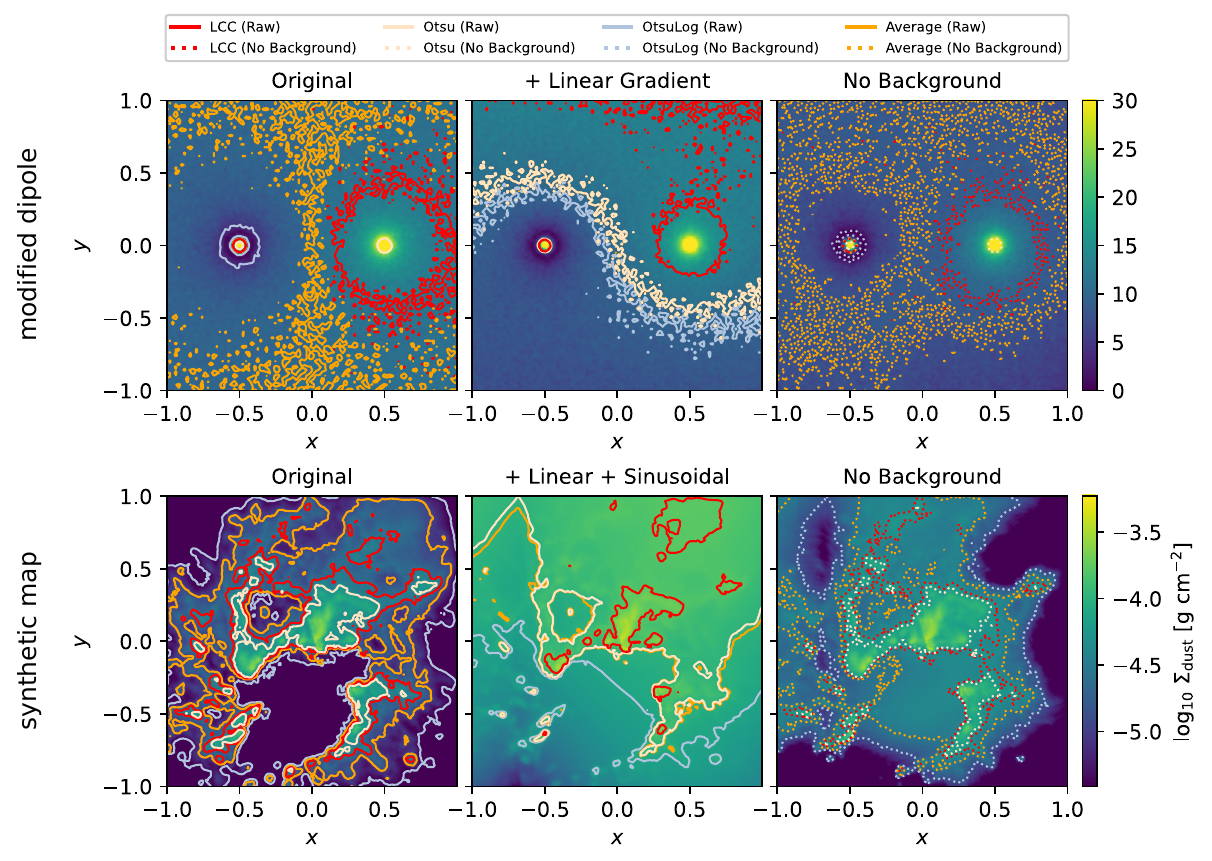}
\caption{Test of the background removal algorithm on a modified dipole function (\crefalt{eq:modified_dipole}; top panels) and a synthetic map of a star-forming region (bottom panels). The panels on the left show the original, unperturbed images for each case. In the middle panels, images are perturbed from their original state with a known sinusoidal and/or linear gradient (see \namecrefs{eq:modified_dipole} \ref{eq:modified_dipole} and \ref{eq:sph_perturbation}). The panels on the right show the result after applying the background removal to the perturbed images in an attempt to recover the original images on the left. While the algorithm effectively restores the morphology of high-density regions, many low-density features tend to be lost.}
\label{fig:test_nobackground}
\end{figure*}

While the average threshold contour utilizes the \href{https://docs.scipy.org/doc/scipy/reference/ndimage.html}{\texttt{scipy.ndimage}} package \citep{Virtanen/etal/2020}, we pair it with a tunable percentile threshold that can be adjusted based on the image's visual or gradient features. Setting the percentile threshold close to zero results in the lowest contours, whereas a threshold near 100 yields the highest contours. For example, in the bottom left panel of \cref{fig:test_nobackground}, the Otsu and OtsuLog threshold contours correspond to percentiles of $84.3$ and $5.6$, respectively. The benefit of using the average threshold contour over manual level selection lies in its ability to establish a direct link to the physical gradients present in the image. This systematic approach aids application across various cases, enhancing both consistency and interpretability.

\begin{table*}
    \centering
    \caption{Values for the contour levels shown in \cref{fig:test_nobackground}. Background removal algorithms often vary in their compensation for contamination across different regions, affecting contour values even when the original contour shapes are restored. In the case of the perturbed modified dipole data, background removal resulted in reduced contour values, whereas the opposite effect was observed in the synthetic star-forming maps.}
    \label{tab:contour_values}
    \sisetup{
        round-mode              = places,
        round-precision         = 3,
        scientific-notation     = true,
        output-exponent-marker  = \text{e},
        table-align-text-pre    = false,
        table-align-text-post   = false,
    }
    \begin{tabular*}{\textwidth}{
        l
        l
        @{\extracolsep{\fill}}
        S[table-format=1.3e-1]
        S[table-format=1.3e-1]
        S[table-format=1.3e-1]
        S[table-format=1.3e-1]
    }
        \toprule
           &   Data             & {LCC}                     & {Otsu}                   & {OtsuLog}                & {Average}                \\
        \midrule
       \multirow{3}{*}{\rotatebox[origin=c]{90}{modified} \rotatebox[origin=c]{90}{dipole}} &   Original         & 11.91798034387492         & 29.533118256477707       & 5.191567721845911        & 10.504917718576063       \\
        &   Perturbed        & 13.997242813584458        & 10.675029229439584       & 9.82955663402621         & 10.37408663575168        \\
        &   No Background    & 9.673782267869514         & 27.95009419695019        & 1.47845274727813         & 7.737901312883389        \\
        \midrule
        \multirow{3}{*}{\rotatebox[origin=c]{90}{synthetic} \rotatebox[origin=c]{90}{map}} &   Original         & 2.5622711198720822e-05    & 4.410976344695388e-05    & 3.4543278962667862e-06   & 1.073520773407142e-05    \\
        &   Perturbed        & 0.00015920654645858027    & 0.00010108460019405182   & 7.279582364208205e-05    & 9.820989894417076e-05    \\
        &   No Background    & 4.6054094535230674e-05    & 6.117563039907938e-05    & 1.339677668895176e-05    & 2.8903253710923785e-05   \\
        \bottomrule
    \end{tabular*}
\end{table*}

To test the background removal feature in the code, we constructed a non-negative, modified dipole with noise and a linear gradient:
\begin{equation}
    \Sigma(x, y) = \left| \frac{1}{\sqrt{(x - \frac{1}{2})^2 + y^2}} - \frac{1}{\sqrt{(x + \frac{1}{2})^2 + y^2}} + \xi \right|
    \label{eq:modified_dipole}
\end{equation}
where $\xi = 10 + 0.5 \cdot \mathcal{N}(0, 1) + 3 y$ and $\mathcal{N}(0, 1)$ represents a standard normal random variable. We chose this function because of its prominent peaks, a distinct background with a local minimum that dips below this background, noise, and a linear gradient to simulate foreground/background contamination. The top three panels of \cref{fig:test_nobackground} illustrate how the remove background feature can potentially account for non-constant background contamination. The left and middle panels show the image and contours that result before/after adding the linear gradient, respectively. This contamination significantly skews the contours of the image, making it difficult to estimate background levels. The LCC is also shifted to higher levels because the gradient opens many of the lower contour levels. The right panel shows the result after applying a `\texttt{square}' background filter with size 70 using the `\texttt{minimum}' pad mode. The contours are in large part restored to their original shape; however, as is often the case with background removal algorithms, over compensation has caused a noticeable drop in signal strength of the low-density regions. The values for each contour are given in \cref{tab:contour_values}.

\begin{figure*}  
\plotone{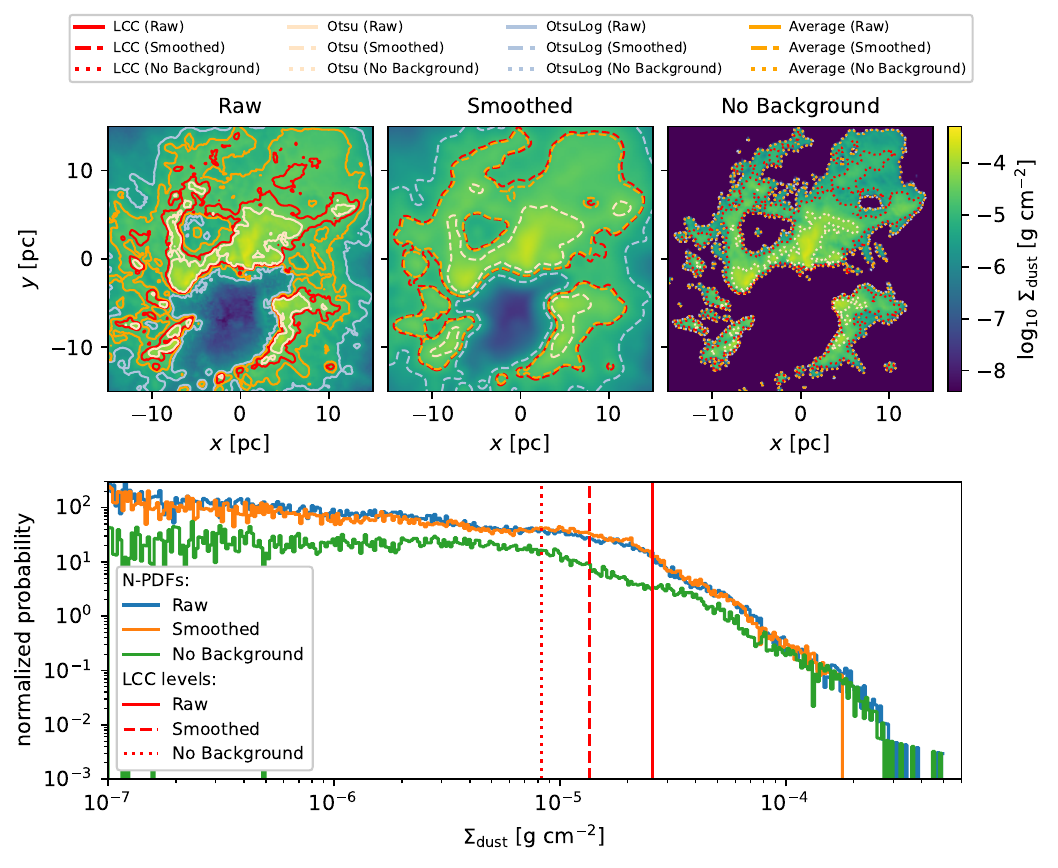}
\caption{{\bf Top}: Illustration of the effects of image processing on the the four contours calculated by the code: unprocessed (left), smoothing (middle), and background removal (right). {\bf Bottom}: Normalized N-PDF for each panel in the top row, with vertical lines indicating the location of their respective LCCs. Smoothing primarily affects high-density signals while background removal affects low-density signals.}
\label{fig:lcc_comparison}
\end{figure*} 

The example above serves as a proof of concept but contrasts sharply with the intricate column density maps typical of star-forming regions. In the bottom three panels of \cref{fig:test_nobackground}, we test how the background removal works on a realistic synthetic map of a star-forming region. Here, we introduce a sinusoidal perturbation in the $x$ direction and a linear gradient in the $y$ direction:
\begin{align}
    \xi_x &= \left|\frac{\max(\Sigma_{\text{raw}})}{10} \cdot \sin\left[\left(\frac{x}{3} + 2\right)^2\right]\right|,
\\
    \xi_y &= \frac{y - y_{\min}}{y_{\max} - y_{\min}} \cdot \frac{\max(\Sigma_{\text{raw}})}{5},
\\
    \Sigma(x, y) &= \Sigma_{\text{raw}}(x, y) + \xi_x(x) + \xi_y(y),
    \label{eq:sph_perturbation}
\end{align}
where $y_{\min}$ and $y_{\max}$ denote the minimum and maximum extents of the image, respectively, and $\Sigma_{\text{raw}}$ represents the unperturbed image. In the middle panel, we observe similar skewed contours as in the modified dipole case, with enhanced signals in the upper and right sections. Applying the background removal algorithm in the panel on the right, we utilize the '\texttt{symmetric}' padding mode with a `\texttt{square}' background filter of size 40. Comparing this with the original unperturbed image on the left, artifacts from the perturbation persist in low-density regions. This is particularly evident in the OtsuLog and Average threshold contours, which are sensitive to low-density signals. Notably, the void at the bottom center remains partially filled, and a vertical line of under-dense material near $x=-2/3$, corresponding to the minimum of the sinusoidal perturbation, can still be seen. However, apart from some missing or altered islands of material, the LCC and Otsu threshold contours closely resemble their original counterparts. This suggests that the morphology of high-density regions has largely been restored, albeit with a systematic shift in contour values compared to the original (see \cref{tab:contour_values}).

The examples shown in \cref{fig:test_nobackground} represent idealized scenarios where we know the underlying structure before applying any background filters. It is crucial to understand that different choices of filters and padding modes can lead to markedly different outcomes. We strongly recommend that users first develop a thorough understanding of how various filters and modes affect the results before applying the algorithm to real data. Consequently, this tool is best employed on a case-by-case basis, with a strong reliance on CAT's interactive plotting routine to fine-tune parameters and achieve optimal results.

\section{Application}
\label{sec:application}

The testing performed in the previous section has already demonstrated some of the functionality of CAT, particularly regarding background removal. Here we focus more particularly on the LCC and how the smoothing, background removal, and masking can affect the resulting N-PDF.

The top three panels in \cref{fig:lcc_comparison} display the contours calculated for the same synthetic map shown earlier: without processing (left panel), after smoothing the image with a Gaussian filter with $\sigma = 3$ (middle panel), and after applying a `\texttt{disk}' background filter of size 150 and the pad mode set to `\texttt{reflect}' (right panel). In the raw image, numerous small clumps are located near or extend off the edge of the map, causing the LCC to shift to higher values -- almost reaching the level of the linear Otsu contour. Typically, the Otsu contour is the highest of the four contours unless the average threshold is specifically adjusted to target high-density regions. Smoothing washes out many of these small clumps, allowing the LCC to shift to lower values. In this case, the LCC is nearly equal to the average contour, which we have calibrated using the lower 20$^{\rm th}$ percentile of the image's gradient magnitudes. Background removal primarily affects the very low-density regions but can also eliminate some small-scale features of the map, resulting in a slight expansion of the LCC. However, the most dramatic shifts in the no-background panel are in the Otsu and OtsuLog contours, which both shift upwards due to the map's new minimum value. Together, these observations highlight the impact that image processing techniques can have on contour calculations.

\begin{figure*}  
\plotone{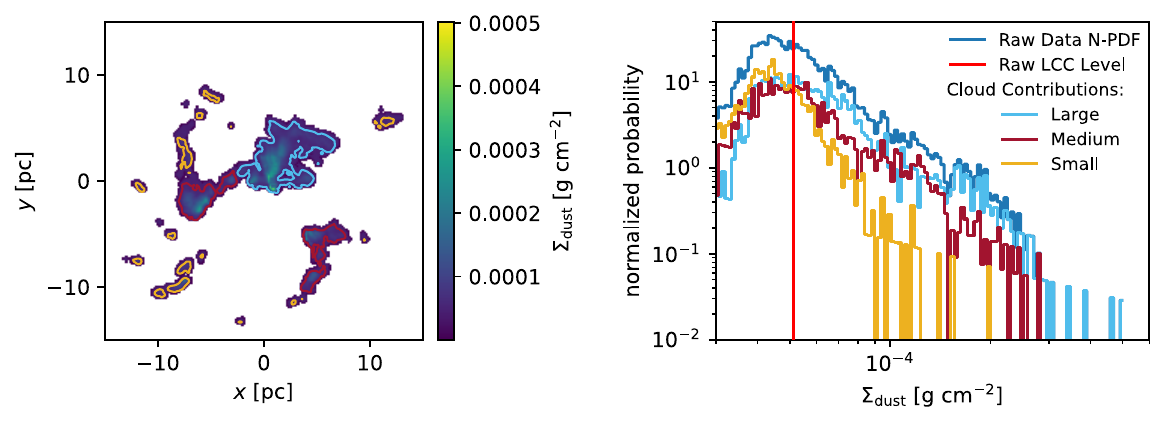}
\caption{Example of how double-masking can be used to disentangle individual cloud contributions to the N-PDF of a large cloud system. {\bf Left}: Region selected by the first Otsu contour mask. The colored lines collectively represent the LCC within this masked region, but are color-coded by cloud group to identify regions manually selected using the second mask. {\bf Right}: N-PDF for the combined region as well as for the individual clouds. The slope of the combined N-PDF appears linear despite the individual contributions being slightly stepped.}
\label{fig:lcc_contour_mask}
\end{figure*} 

The bottom panel of \cref{fig:lcc_comparison} shows the normalized N-PDF for each panel in the top row, with vertical lines indicating the location of their respective LCCs. As expected, smoothing washes out the highest density regions (i.e., $\gtrsim 2 \times 10^{-4} \; \rm g \; cm^{-2}$) but surprisingly has almost no effect on the rest of the N-PDF. This, coupled with the fact that the LCC has now shifted below a clear break in the N-PDF slope, could have important implications for interpreting N-PDFs. In contrast, background removal had minimal impact on the highest density regions but stretched the line profile horizontally at lower densities, accentuating previously minor fluctuations. These fluctuations in slope, still well above the LCC, may carry physical significance and only become visible when the interfering background is removed. A full analysis is beyond the scope of this paper, but we hope the new tools provided by CAT will offer fresh insights into this field.

As a final demonstration, we showcase how two of CAT's unique masking capabilities can be used to obtain individual cloud contributions in the N-PDF of a large cloud system like the one in the synthetic map. First, in an interactive session, we mask the data using the linear Otsu contour, which isolates just the high-density regions of the map. We then save the masked image (e.g. by copying and running the code provided by the `Toggle Info Display' button) and use this as the input image for a second instance of CAT. This allows us to apply a second mask over the original mask. This time, we choose to draw the mask interactively and erase everything but the largest cloud in the center of the image. We save the image of the largest cloud and repeat the previous step to obtain similar images for the medium- and small-sized clouds. Finally, we calculate the N-PDF for all three of these images and normalise them using the first masked image containing all three cloud types.

\Cref{fig:lcc_contour_mask} summarizes the results from the double-masking procedure described above. The left panel shows the regions selected by the first Otsu contour mask. The colored lines collectively represent the LCC within this masked region, but are color-coded by cloud group to identify regions manually selected using the second mask. The right panel displays the N-PDF for each cloud group and the combined N-PDF, with the vertical line marking the LCC location from the left panel. All three cloud groups peak near the LCC, which is expected given that the cloud limbs have the same magnitude and each cloud group covers roughly the same area. At higher densities, the contribution from the small clouds drops off steeply, while the medium and large clouds show more of a stepped profile, similar to the no-background profile in \cref{fig:lcc_comparison}. Despite the uneven profiles in the individual N-PDFs, the combined N-PDF appears surprisingly linear on a log-log scale. Further investigation is required beyond the scope of this paper; nevertheless, this example demonstrates how CAT can assist in analyzing large cloud systems. It additionally shows how to access tool combinations not accounted for in the code itself (e.g. double masking or smoothing of an image where the background has already been removed).

\section{Summary}
\label{sec:summary}

The Contour Analysis Tool (CAT) presented in this paper provides a robust method for identifying and analyzing key structural elements in column density maps using various contouring techniques. These techniques include the lowest-closed contour (LCC), commonly used in N-PDF studies of star-forming regions; linear and logarithmic Otsu thresholding, often found in image processing software; and an average threshold contour that can be tuned based on the gradient features in an image. Contours can assist in identifying lower foreground limits and/or upper background limits for structures at different densities. CAT also offers additional functions for smoothing, background removal, and masking, each with multiple methods and parameters to accommodate a wide range of data and usage scenarios.

One of CAT's most important features is its extensive interactive interface, which enables users to promptly observe the impacts of various methods and parameters. This interface also generates executable code that can be used to reproduce results programmatically at a later time or in a different cell or script. As such, CAT requires minimal coding and knowledge of Python, making it easy to use.

We have outlined the options and algorithms implemented in CAT and tested features that we developed ourselves. Using data from a realistic synthetic map of a star-forming region, we demonstrated CAT's functionality and potential usefulness to the community. Future developments include creating a more versatile GUI that can operate in multiple environments and integrating advanced edge detection software to identify structures spanning multiple orders of magnitude. The tool is publicly available on GitHub at \href{https://github.com/markahutch/ContourAnalysisTool}{https://github.com/markahutch/ContourAnalysisTool}.


\pagebreak

\begin{acknowledgments}
\vspace{-2.5em}
\section*{Acknowledgments}
    We thank the two anonymous referees for their thorough and constructive reports, which significantly improved the scope and clarity of this paper. This work was supported by the DFG program ``Closing the Loop - Using Synthetic Observations of Simulated Star-forming Regions to Test Observational Properties" (DFG Project Number: 426714422). C.K. acknowledges support from the Bayerischen Gleichstellungsf{\"o}rderung. The authors would like to acknowledge OpenAI's ChatGPT for providing coding assistance and manuscript preparation. We also thank `\href{https://stackoverflow.com/users/566326/developer}{Developer}' for their solution in the ``\href{https://stackoverflow.com/questions/19849167/find-most-distant-points-in-contour-curve/19849862}{Find most distant points in contour curve}" thread on stackoverflow.com, which inspired our approach to finding the LCC.
\end{acknowledgments}

\bibliography{ref}{}

\begin{thebibliography}{}
\expandafter\ifx\csname natexlab\endcsname\relax\def\natexlab#1{#1}\fi
\providecommand{\url}[1]{\href{#1}{#1}}
\providecommand{\dodoi}[1]{doi:~\href{http://doi.org/#1}{\nolinkurl{#1}}}
\providecommand{\doeprint}[1]{\href{http://ascl.net/#1}{\nolinkurl{http://ascl.net/#1}}}
\providecommand{\doarXiv}[1]{\href{https://arxiv.org/abs/#1}{\nolinkurl{https://arxiv.org/abs/#1}}}

\bibitem[{{Alves} {et~al.}(2017){Alves}, {Lombardi}, \&
  {Lada}}]{Alves/Lombardi/Lada/2017}
{Alves}, J., {Lombardi}, M., \& {Lada}, C.~J. 2017, \aap, 606, L2,
  \dodoi{10.1051/0004-6361/201731436}

\bibitem[{{Ballesteros-Paredes} {et~al.}(2011){Ballesteros-Paredes},
  {V{\'a}zquez-Semadeni}, {Gazol}, {Hartmann}, {Heitsch}, \&
  {Col{\'\i}n}}]{Ballesteros-Paredes/etal/2011}
{Ballesteros-Paredes}, J., {V{\'a}zquez-Semadeni}, E., {Gazol}, A., {et~al.}
  2011, \mnras, 416, 1436, \dodoi{10.1111/j.1365-2966.2011.19141.x}

\bibitem[{{Berry}(2015)}]{Berry/2015}
{Berry}, D.~S. 2015, Astronomy and Computing, 10, 22,
  \dodoi{10.1016/j.ascom.2014.11.004}

\bibitem[{{Bigiel} {et~al.}(2011){Bigiel}, {Leroy}, {Walter}, {Brinks}, {de
  Blok}, {Kramer}, {Rix}, {Schruba}, {Schuster}, {Usero}, \&
  {Wiesemeyer}}]{Bigiel/etal/2011}
{Bigiel}, F., {Leroy}, A.~K., {Walter}, F., {et~al.} 2011, \apjl, 730, L13,
  \dodoi{10.1088/2041-8205/730/2/L13}

\bibitem[{Buades {et~al.}(2011)Buades, Coll, \& Morel}]{Buades/Coll/Morel/2011}
Buades, A., Coll, B., \& Morel, J.-M. 2011, {Image Processing On Line}, 1, 208

\bibitem[{Cleveland \& Devlin(1988)}]{Cleveland/Devlin/1988}
Cleveland, W.~S., \& Devlin, S.~J. 1988, Journal of the American Statistical
  Association, 83, 596, \dodoi{10.1080/01621459.1988.10478639}

\bibitem[{{Colombo} {et~al.}(2015){Colombo}, {Rosolowsky}, {Ginsburg},
  {Duarte-Cabral}, \& {Hughes}}]{Colombo/etal/2015}
{Colombo}, D., {Rosolowsky}, E., {Ginsburg}, A., {Duarte-Cabral}, A., \&
  {Hughes}, A. 2015, \mnras, 454, 2067, \dodoi{10.1093/mnras/stv2063}

\bibitem[{{Dale} {et~al.}(2014){Dale}, {Ngoumou}, {Ercolano}, \&
  {Bonnell}}]{Dale/etal/2014}
{Dale}, J.~E., {Ngoumou}, J., {Ercolano}, B., \& {Bonnell}, I.~A. 2014, \mnras,
  442, 694, \dodoi{10.1093/mnras/stu816}

\bibitem[{Dierckx(1993)}]{Dierckx/1993}
Dierckx, P. 1993, Curve and surface fitting with splines (USA: Oxford
  University Press, Inc.)

\bibitem[{{Federrath} \& {Klessen}(2013)}]{Federrath/Klessen/2013}
{Federrath}, C., \& {Klessen}, R.~S. 2013, \apj, 763, 51,
  \dodoi{10.1088/0004-637X/763/1/51}

\bibitem[{{Federrath} {et~al.}(2016){Federrath}, {Rathborne}, {Longmore},
  {Kruijssen}, {Bally}, {Contreras}, {Crocker}, {Garay}, {Jackson}, {Testi}, \&
  {Walsh}}]{Federrath/etal/2016}
{Federrath}, C., {Rathborne}, J.~M., {Longmore}, S.~N., {et~al.} 2016, \apj,
  832, 143, \dodoi{10.3847/0004-637X/832/2/143}

\bibitem[{{Fujimoto} {et~al.}(2019){Fujimoto}, {Chevance}, {Haydon},
  {Krumholz}, \& {Kruijssen}}]{Fujimoto/etal/2019}
{Fujimoto}, Y., {Chevance}, M., {Haydon}, D.~T., {Krumholz}, M.~R., \&
  {Kruijssen}, J.~M.~D. 2019, \mnras, 487, 1717, \dodoi{10.1093/mnras/stz641}

\bibitem[{Gonzalez \& Woods(2017)}]{Gonzalez/Woods/2017}
Gonzalez, R., \& Woods, R. 2017, {Digital Image Processing Global Edition}
  (Pearson Deutschland), 1024.
\newblock \url{https://elibrary.pearson.de/book/99.150005/9781292223070}

\bibitem[{{Goodman} {et~al.}(2009){Goodman}, {Pineda}, \&
  {Schnee}}]{Goodman/Pineda/Schnee/2009}
{Goodman}, A.~A., {Pineda}, J.~E., \& {Schnee}, S.~L. 2009, \apj, 692, 91,
  \dodoi{10.1088/0004-637X/692/1/91}

\bibitem[{{Gouliermis} {et~al.}(2010){Gouliermis}, {Schmeja}, {Klessen}, {de
  Blok}, \& {Walter}}]{Gouliermis/etal/2010}
{Gouliermis}, D.~A., {Schmeja}, S., {Klessen}, R.~S., {de Blok}, W.~J.~G., \&
  {Walter}, F. 2010, \apj, 725, 1717, \dodoi{10.1088/0004-637X/725/2/1717}

\bibitem[{{Grasha} {et~al.}(2017){Grasha}, {Calzetti}, {Adamo}, {Kim},
  {Elmegreen}, {Gouliermis}, {Dale}, {Fumagalli}, {Grebel}, {Johnson}, {Kahre},
  {Kennicutt}, {Messa}, {Pellerin}, {Ryon}, {Smith}, {Shabani}, {Thilker}, \&
  {Ubeda}}]{Grasha/etal/2017}
{Grasha}, K., {Calzetti}, D., {Adamo}, A., {et~al.} 2017, \apj, 840, 113,
  \dodoi{10.3847/1538-4357/aa6f15}

\bibitem[{{Handa} {et~al.}(1990){Handa}, {Nakai}, {Sofue}, {Hayashi}, \&
  {Fujimoto}}]{Handa/etal/1990}
{Handa}, T., {Nakai}, N., {Sofue}, Y., {Hayashi}, M., \& {Fujimoto}, M. 1990,
  \pasj, 42, 1

\bibitem[{{Harris} {et~al.}(2020){Harris}, {Millman}, {van der Walt},
  {Gommers}, {Virtanen}, {Cournapeau}, {Wieser}, {Taylor}, {Berg}, {Smith},
  {Kern}, {Picus}, {Hoyer}, {van Kerkwijk}, {Brett}, {Haldane}, {del R{\'\i}o},
  {Wiebe}, {Peterson}, {G{\'e}rard-Marchant}, {Sheppard}, {Reddy}, {Weckesser},
  {Abbasi}, {Gohlke}, \& {Oliphant}}]{Harris/etal/2020}
{Harris}, C.~R., {Millman}, K.~J., {van der Walt}, S.~J., {et~al.} 2020, \nat,
  585, 357, \dodoi{10.1038/s41586-020-2649-2}

\bibitem[{{Jaupart} \& {Chabrier}(2020)}]{Jaupart/Chabrier/2020}
{Jaupart}, E., \& {Chabrier}, G. 2020, \apjl, 903, L2,
  \dodoi{10.3847/2041-8213/abbda8}

\bibitem[{{Jiao} {et~al.}(2022){Jiao}, {Wu}, {Ruan}, {Lin}, {Tsai}, \&
  {Feng}}]{Jiao/etal/2022}
{Jiao}, S., {Wu}, J., {Ruan}, H., {et~al.} 2022, Research in Astronomy and
  Astrophysics, 22, 075016, \dodoi{10.1088/1674-4527/ac6850}

\bibitem[{{Kainulainen} {et~al.}(2011){Kainulainen}, {Beuther}, {Banerjee},
  {Federrath}, \& {Henning}}]{Kainulainen/etal/2011}
{Kainulainen}, J., {Beuther}, H., {Banerjee}, R., {Federrath}, C., \&
  {Henning}, T. 2011, \aap, 530, A64, \dodoi{10.1051/0004-6361/201016383}

\bibitem[{{Kainulainen} {et~al.}(2009){Kainulainen}, {Beuther}, {Henning}, \&
  {Plume}}]{Kainulainen/etal/2009}
{Kainulainen}, J., {Beuther}, H., {Henning}, T., \& {Plume}, R. 2009, \aap,
  508, L35, \dodoi{10.1051/0004-6361/200913605}

\bibitem[{{Kainulainen} {et~al.}(2013){Kainulainen}, {Federrath}, \&
  {Henning}}]{Kainulainen/Federrath/Henning/2013}
{Kainulainen}, J., {Federrath}, C., \& {Henning}, T. 2013, \aap, 553, L8,
  \dodoi{10.1051/0004-6361/201321431}

\bibitem[{{Kennicutt}(1998)}]{Kennicutt/1998}
{Kennicutt}, Robert~C., J. 1998, \apj, 498, 541, \dodoi{10.1086/305588}

\bibitem[{{Khullar} {et~al.}(2021){Khullar}, {Federrath}, {Krumholz}, \&
  {Matzner}}]{Khullar/etal/2021}
{Khullar}, S., {Federrath}, C., {Krumholz}, M.~R., \& {Matzner}, C.~D. 2021,
  \mnras, 507, 4335, \dodoi{10.1093/mnras/stab1914}

\bibitem[{{Koch} \& {Rosolowsky}(2015)}]{Koch/Rosolowsky/2015}
{Koch}, E.~W., \& {Rosolowsky}, E.~W. 2015, \mnras, 452, 3435,
  \dodoi{10.1093/mnras/stv1521}

\bibitem[{{Koepferl} \& {Robitaille}(2017)}]{Koepferl/Robitaille/2017}
{Koepferl}, C.~M., \& {Robitaille}, T.~P. 2017, \apj, 849, 3,
  \dodoi{10.3847/1538-4357/aa8666}

\bibitem[{{Koepferl} {et~al.}(2017{\natexlab{a}}){Koepferl}, {Robitaille}, \&
  {Dale}}]{Koepferl/Robitaille/Dale/2017a}
{Koepferl}, C.~M., {Robitaille}, T.~P., \& {Dale}, J.~E. 2017{\natexlab{a}},
  \apj, 849, 1, \dodoi{10.3847/1538-4357/849/1/1}

\bibitem[{{Koepferl} {et~al.}(2017{\natexlab{b}}){Koepferl}, {Robitaille},
  {Dale}, \& {Biscani}}]{Koepferl/etal/2017}
{Koepferl}, C.~M., {Robitaille}, T.~P., {Dale}, J.~E., \& {Biscani}, F.
  2017{\natexlab{b}}, \apjs, 233, 1, \dodoi{10.3847/1538-4365/233/1/1}

\bibitem[{{K{\"o}rtgen} {et~al.}(2019){K{\"o}rtgen}, {Federrath}, \&
  {Banerjee}}]{Kortgen/Federrath/Banerjee/2019}
{K{\"o}rtgen}, B., {Federrath}, C., \& {Banerjee}, R. 2019, \mnras, 482, 5233,
  \dodoi{10.1093/mnras/sty3071}

\bibitem[{{Lada} {et~al.}(2010){Lada}, {Lombardi}, \&
  {Alves}}]{Lada/Lombardi/Alves/2010}
{Lada}, C.~J., {Lombardi}, M., \& {Alves}, J.~F. 2010, \apj, 724, 687.
\newblock \doarXiv{1009.2985}

\bibitem[{{Lewis} {et~al.}(2022){Lewis}, {Lada}, \&
  {Dame}}]{Lewis/Lada/Dame/2022}
{Lewis}, J.~A., {Lada}, C.~J., \& {Dame}, T.~M. 2022, \apj, 931, 9,
  \dodoi{10.3847/1538-4357/ac5d58}

\bibitem[{{Lombardi} {et~al.}(2015){Lombardi}, {Alves}, \&
  {Lada}}]{Lombardi/Alves/Lada/2015}
{Lombardi}, M., {Alves}, J., \& {Lada}, C.~J. 2015, \aap, 576, L1,
  \dodoi{10.1051/0004-6361/201525650}

\bibitem[{{Ma} {et~al.}(2021){Ma}, {Wang}, {Li}, {Lin}, {Sun}, \&
  {Yang}}]{Ma/etal/2021}
{Ma}, Y., {Wang}, H., {Li}, C., {et~al.} 2021, \apjs, 254, 3,
  \dodoi{10.3847/1538-4365/abe85c}

\bibitem[{{Ma} {et~al.}(2022){Ma}, {Wang}, {Zhang}, {Wang}, {Zhang}, {Liu},
  {Li}, {Zheng}, {Yuan}, \& {Yang}}]{Ma/etal/2022}
{Ma}, Y., {Wang}, H., {Zhang}, M., {et~al.} 2022, \apjs, 262, 16,
  \dodoi{10.3847/1538-4365/ac7797}

\bibitem[{{Ma} {et~al.}(2020){Ma}, {Wang}, {Li}, \& {Yang}}]{Ma/etal/2020}
{Ma}, Y.-H., {Wang}, H.-C., {Li}, C., \& {Yang}, J. 2020, Research in Astronomy
  and Astrophysics, 20, 060, \dodoi{10.1088/1674-4527/20/4/60}

\bibitem[{{Maeda} {et~al.}(2023){Maeda}, {Egusa}, {Ohta}, {Fujimoto}, \&
  {Habe}}]{Maeda/etal/2023}
{Maeda}, F., {Egusa}, F., {Ohta}, K., {Fujimoto}, Y., \& {Habe}, A. 2023, \apj,
  943, 7, \dodoi{10.3847/1538-4357/aca664}

\bibitem[{{Mattern} {et~al.}(2024){Mattern}, {Andr{\'e}}, {Zavagno}, {Russeil},
  {Roussel}, {Peretto}, {Schuller}, {Shimajiri}, {Di Francesco}, {Arzoumanian},
  {Rev{\'e}ret}, \& {De Breuck}}]{Mattern/etal/2024}
{Mattern}, M., {Andr{\'e}}, P., {Zavagno}, A., {et~al.} 2024, arXiv e-prints,
  arXiv:2405.15713, \dodoi{10.48550/arXiv.2405.15713}

\bibitem[{{Molinari} {et~al.}(2011){Molinari}, {Schisano}, {Faustini},
  {Pestalozzi}, {di Giorgio}, \& {Liu}}]{Molinari/etal/2011}
{Molinari}, S., {Schisano}, E., {Faustini}, F., {et~al.} 2011, \aap, 530, A133,
  \dodoi{10.1051/0004-6361/201014752}

\bibitem[{{Momose} {et~al.}(2010){Momose}, {Okumura}, {Koda}, \&
  {Sawada}}]{Momose/etal/2010}
{Momose}, R., {Okumura}, S.~K., {Koda}, J., \& {Sawada}, T. 2010, \apj, 721,
  383, \dodoi{10.1088/0004-637X/721/1/383}

\bibitem[{{Murase} {et~al.}(2022){Murase}, {Handa}, {Hirata}, {Omodaka},
  {Nakano}, {Sunada}, {Shimajiri}, \& {Nishi}}]{Murase/etal/2022}
{Murase}, T., {Handa}, T., {Hirata}, Y., {et~al.} 2022, \mnras, 510, 1106,
  \dodoi{10.1093/mnras/stab3472}

\bibitem[{{Neyrinck}(2008)}]{Neyrinck/2008}
{Neyrinck}, M.~C. 2008, \mnras, 386, 2101,
  \dodoi{10.1111/j.1365-2966.2008.13180.x}

\bibitem[{{Ossenkopf-Okada} {et~al.}(2016){Ossenkopf-Okada}, {Csengeri},
  {Schneider}, {Federrath}, \& {Klessen}}]{Ossenkopf-Okada/etal/2016}
{Ossenkopf-Okada}, V., {Csengeri}, T., {Schneider}, N., {Federrath}, C., \&
  {Klessen}, R.~S. 2016, \aap, 590, A104, \dodoi{10.1051/0004-6361/201628095}

\bibitem[{{Ostriker} {et~al.}(2001){Ostriker}, {Stone}, \&
  {Gammie}}]{Ostriker/Stone/Gammie/2001}
{Ostriker}, E.~C., {Stone}, J.~M., \& {Gammie}, C.~F. 2001, \apj, 546, 980,
  \dodoi{10.1086/318290}

\bibitem[{Otsu(1979)}]{Otsu/1979}
Otsu, N. 1979, IEEE Transactions on Systems, Man, and Cybernetics, 9, 62,
  \dodoi{10.1109/TSMC.1979.4310076}

\bibitem[{Perona \& Malik(1990)}]{Perona/Malik/1990}
Perona, P., \& Malik, J. 1990, IEEE Transactions on Pattern Analysis and
  Machine Intelligence, 12, 629, \dodoi{10.1109/34.56205}

\bibitem[{{Pokhrel} {et~al.}(2016){Pokhrel}, {Gutermuth}, {Ali}, {Megeath},
  {Pipher}, {Myers}, {Fischer}, {Henning}, {Wolk}, {Allen}, \&
  {Tobin}}]{Pokhrel/etal/2016}
{Pokhrel}, R., {Gutermuth}, R., {Ali}, B., {et~al.} 2016, \mnras, 461, 22,
  \dodoi{10.1093/mnras/stw1303}

\bibitem[{{Rosolowsky} \& {Leroy}(2006)}]{Rosolowsky/Leroy/2006}
{Rosolowsky}, E., \& {Leroy}, A. 2006, \pasp, 118, 590, \dodoi{10.1086/502982}

\bibitem[{{Rosolowsky} {et~al.}(2008){Rosolowsky}, {Pineda}, {Kauffmann}, \&
  {Goodman}}]{Rosolowsky/etal/2008}
{Rosolowsky}, E.~W., {Pineda}, J.~E., {Kauffmann}, J., \& {Goodman}, A.~A.
  2008, \apj, 679, 1338, \dodoi{10.1086/587685}

\bibitem[{{Rudin} {et~al.}(1992){Rudin}, {Osher}, \&
  {Fatemi}}]{Rudin/Osher/Fatemi/1992}
{Rudin}, L.~I., {Osher}, S., \& {Fatemi}, E. 1992, Physica D Nonlinear
  Phenomena, 60, 259, \dodoi{10.1016/0167-2789(92)90242-F}

\bibitem[{Russ(2011)}]{Russ/2011}
Russ, J.~C. 2011, The Image Processing Handbook, 6th edn. (CRC Press),
  \dodoi{10.1201/b10720}

\bibitem[{Schafer(2011)}]{Schafer/2011}
Schafer, R.~W. 2011, IEEE Signal Processing Magazine, 28, 111,
  \dodoi{10.1109/MSP.2011.941097}

\bibitem[{{Schneider} {et~al.}(2013){Schneider}, {Andr{\'e}}, {K{\"o}nyves},
  {Bontemps}, {Motte}, {Federrath}, {Ward-Thompson}, {Arzoumanian},
  {Benedettini}, {Bressert}, {Didelon}, {Di Francesco}, {Griffin}, {Hennemann},
  {Hill}, {Palmeirim}, {Pezzuto}, {Peretto}, {Roy}, {Rygl}, {Spinoglio}, \&
  {White}}]{Schneider/etal/2013}
{Schneider}, N., {Andr{\'e}}, P., {K{\"o}nyves}, V., {et~al.} 2013, \apjl, 766,
  L17, \dodoi{10.1088/2041-8205/766/2/L17}

\bibitem[{{Schneider} {et~al.}(2015{\natexlab{a}}){Schneider}, {Bontemps},
  {Girichidis}, {Rayner}, {Motte}, {Andr{\'e}}, {Russeil}, {Abergel},
  {Anderson}, {Arzoumanian}, {Benedettini}, {Csengeri}, {Didelon}, {di},
  {Griffin}, {Hill}, {Klessen}, {Ossenkopf}, {Pezzuto}, {Rivera-Ingraham},
  {Spinoglio}, {Tremblin}, \& {Zavagno}}]{Schneider/etal/2015a}
{Schneider}, N., {Bontemps}, S., {Girichidis}, P., {et~al.} 2015{\natexlab{a}},
  \mnras, 453, L41, \dodoi{10.1093/mnrasl/slv101}

\bibitem[{{Schneider} {et~al.}(2015{\natexlab{b}}){Schneider}, {Ossenkopf},
  {Csengeri}, {Klessen}, {Federrath}, {Tremblin}, {Girichidis}, {Bontemps}, \&
  {Andr{\'e}}}]{Schneider/etal/2015b}
{Schneider}, N., {Ossenkopf}, V., {Csengeri}, T., {et~al.} 2015{\natexlab{b}},
  \aap, 575, A79, \dodoi{10.1051/0004-6361/201423569}

\bibitem[{{Schneider} {et~al.}(2016){Schneider}, {Bontemps}, {Motte},
  {Ossenkopf}, {Klessen}, {Simon}, {Fechtenbaum}, {Herpin}, {Tremblin},
  {Csengeri}, {Myers}, {Hill}, {Cunningham}, \&
  {Federrath}}]{Schneider/etal/2016a}
{Schneider}, N., {Bontemps}, S., {Motte}, F., {et~al.} 2016, \aap, 587, A74,
  \dodoi{10.1051/0004-6361/201527144}

\bibitem[{{Schneider} {et~al.}(2022){Schneider}, {Ossenkopf-Okada}, {Clarke},
  {Klessen}, {Kabanovic}, {Veltchev}, {Bontemps}, {Dib}, {Csengeri},
  {Federrath}, {Di Francesco}, {Motte}, {Andr{\'e}}, {Arzoumanian}, {Beattie},
  {Bonne}, {Didelon}, {Elia}, {K{\"o}nyves}, {Kritsuk}, {Ladjelate}, {Myers},
  {Pezzuto}, {Robitaille}, {Roy}, {Seifried}, {Simon}, {Soler}, \&
  {Ward-Thompson}}]{Schneider/etal/2022}
{Schneider}, N., {Ossenkopf-Okada}, V., {Clarke}, S., {et~al.} 2022, \aap, 666,
  A165, \dodoi{10.1051/0004-6361/202039610}

\bibitem[{Smith {et~al.}(1997)}]{Smith/1997}
Smith, S., {et~al.} 1997, The scientist and engineer's guide to digital signal
  processing (California Technical Pub. San Diego)

\bibitem[{{Sousbie}(2011)}]{Sousbie/2011}
{Sousbie}, T. 2011, \mnras, 414, 350, \dodoi{10.1111/j.1365-2966.2011.18394.x}

\bibitem[{{Stutzki} \& {Guesten}(1990)}]{Stutzki/Guesten/1990}
{Stutzki}, J., \& {Guesten}, R. 1990, \apj, 356, 513, \dodoi{10.1086/168859}

\bibitem[{{Sutter} {et~al.}(2015){Sutter}, {Lavaux}, {Hamaus}, {Pisani},
  {Wandelt}, {Warren}, {Villaescusa-Navarro}, {Zivick}, {Mao}, \&
  {Thompson}}]{Sutter/etal/2015}
{Sutter}, P.~M., {Lavaux}, G., {Hamaus}, N., {et~al.} 2015, Astronomy and
  Computing, 9, 1, \dodoi{10.1016/j.ascom.2014.10.002}

\bibitem[{Tomasi \& Manduchi(1998)}]{Tomasi/Manduchi/1998}
Tomasi, C., \& Manduchi, R. 1998, in Sixth International Conference on Computer
  Vision (IEEE Cat. No.98CH36271), 839--846, \dodoi{10.1109/ICCV.1998.710815}

\bibitem[{van~der Walt {et~al.}(2014)van~der Walt, Sch{\"o}nberger,
  Nunez-Iglesias, Boulogne, Warner, Yager, Gouillart, Yu, \& the scikit-image
  contributors}]{vanderWalt/etal/2014}
van~der Walt, S., Sch{\"o}nberger, J.~L., Nunez-Iglesias, J., {et~al.} 2014,
  PeerJ, 2, e453, \dodoi{10.7717/peerj.453}

\bibitem[{{Vazquez-Semadeni}(1994)}]{Vazquez-Semadeni/1994}
{Vazquez-Semadeni}, E. 1994, \apj, 423, 681, \dodoi{10.1086/173847}

\bibitem[{{V{\'a}zquez-Semadeni} \&
  {Garc{\'\i}a}(2001)}]{Vazquez-Semadeni/Garcia/2001}
{V{\'a}zquez-Semadeni}, E., \& {Garc{\'\i}a}, N. 2001, \apj, 557, 727,
  \dodoi{10.1086/321688}

\bibitem[{{Villanueva} {et~al.}(2022){Villanueva}, {Bolatto}, {Vogel}, {Brown},
  {Wilson}, {Zabel}, {Ellison}, {Stevens}, {Jim{\'e}nez Donaire}, {Spekkens},
  {Tharp}, {Davis}, {Parker}, {Roberts}, {Basra}, {Boselli}, {Catinella},
  {Chung}, {Cortese}, {Lee}, \& {Watts}}]{Villanueva/etal/2022}
{Villanueva}, V., {Bolatto}, A.~D., {Vogel}, S., {et~al.} 2022, \apj, 940, 176,
  \dodoi{10.3847/1538-4357/ac9d3c}

\bibitem[{{Virtanen} {et~al.}(2020){Virtanen}, {Gommers}, {Oliphant},
  {Haberland}, {Reddy}, {Cournapeau}, {Burovski}, {Peterson}, {Weckesser},
  {Bright}, {van der Walt}, {Brett}, {Wilson}, {Millman}, {Mayorov}, {Nelson},
  {Jones}, {Kern}, {Larson}, {Carey}, {Polat}, {Feng}, {Moore}, {VanderPlas},
  {Laxalde}, {Perktold}, {Cimrman}, {Henriksen}, {Quintero}, {Harris},
  {Archibald}, {Ribeiro}, {Pedregosa}, {van Mulbregt}, \& {SciPy 1. 0
  Contributors}}]{Virtanen/etal/2020}
{Virtanen}, P., {Gommers}, R., {Oliphant}, T.~E., {et~al.} 2020, Nature
  Methods, 17, 261, \dodoi{10.1038/s41592-019-0686-2}

\bibitem[{{Wang} {et~al.}(2020){Wang}, {Beuther}, {Schneider}, {Meidt}, {Linz},
  {Ragan}, {Zucker}, {Battersby}, {Soler}, {Schinnerer}, {Bigiel}, {Colombo},
  \& {Henning}}]{Wang/etal/2020}
{Wang}, Y., {Beuther}, H., {Schneider}, N., {et~al.} 2020, \aap, 641, A53,
  \dodoi{10.1051/0004-6361/202037928}

\bibitem[{Wiener(1949)}]{Wiener/1949}
Wiener, N. 1949, Extrapolation, interpolation, and smoothing of stationary time
  series: with engineering applications (The MIT press)

\bibitem[{{Williams} {et~al.}(1994){Williams}, {de Geus}, \&
  {Blitz}}]{Williams/deGeus/Blitz/1994}
{Williams}, J.~P., {de Geus}, E.~J., \& {Blitz}, L. 1994, \apj, 428, 693,
  \dodoi{10.1086/174279}

\bibitem[{{Zhang} {et~al.}(2019){Zhang}, {Kainulainen}, {Mattern}, {Fang}, \&
  {Henning}}]{Zhang/etal/2019}
{Zhang}, M., {Kainulainen}, J., {Mattern}, M., {Fang}, M., \& {Henning}, T.
  2019, \aap, 622, A52, \dodoi{10.1051/0004-6361/201732400}

\end{thebibliography}
\bibliographystyle{aasjournal}

\end{document}